          [2001/04/25 1.1 (PWD)]
\documentclass[a4paper,twocolumn]{esapub}

\usepackage{natbib}

\usepackage{graphicx}
\usepackage{txfonts}
\usepackage{times}

\title{{\it INTEGRAL\/} observations of Cygnus X-3}

\author{Linnea Hjalmarsdotter}
\affil{Observatory, P.O. Box 14, 00014 University of Helsinki, Finland}
\author{A. A. Zdziarski}
\affil{ Centrum Astronomiczne im. M. Kopernika, Bartycka 18, 00-716 Warsawa, Poland}
\author[3,4]{A. Paizis}
\affil[3]{{\it INTEGRAL} Science Data Center, Chemin d'\'Ecogia 16, CH-1290 Versoix, Switzerland}
\affil[4]{CNR-IASF, Sezione di Milano, Via Bassini 15, 20133 Milano, Italy}
\author[4,5]{V. Beckmann}
\affil[4]{NASA Goddard Space Flight Center, Code 661, Greenbelt, MD 20771, USA}
\affil[5]{Joint Center for Astrophysics, Department of Physics, University of Maryland, Baltimore County, MD 21250, USA}
\author[1,3]{O. Vilhu}

\newcommand{\efe}{E_{{\rm K}\alpha}}
\newcommand{\ffe}{F_{{\rm K}\alpha}}
\newcommand{\sfe}{\sigma_{{\rm K}\alpha}}
\newcommand{\ee}{$e^\pm$}
\newcommand{\g}{$\gamma$}
\newcommand{\lh}{\ell_{\rm h}}
\newcommand{\ls}{\ell_{\rm s}}

\newcommand{\xte}{{\it RXTE}}
\newcommand{\rxte}{{\it RXTE}}

\newcommand{\asm}{{\it RXTE}/ASM}
\newcommand{\integral}{{\it INTEGRAL}}

\begin{document}

\keywords{gamma rays: observations -- radiation mechanisms: non-thermal -- stars: individual: Cyg X-3 -- X-rays: binaries -- X-rays: general -- X-rays: stars}

\maketitle

\begin{abstract}
The peculiar X-ray binary Cygnus X-3 has been observed on several occasions with the X/\g-ray instruments on board \integral. We have collected data from available public and Galactic Plane Scan observations between December 2002 and December 2003 and summed them together into two broad-band spectra, representing different physical spectral states of the source. We have fitted the two spectra with models including Comptonization and Compton reflection, similar to those found for black-hole binaries at high accretion rates. 
\end{abstract}

\section{Introduction}

Cyg X-3 is an enigmatic X-ray binary that ever since its discovery 38 
years ago, in 1966 (Giacconi et al.\ 1967), has evaded simple classification. 
Its 4.8 hr orbital period is typical of a low-mass system, but infrared 
observations have shown the donor to be a high-mass Wolf-Rayet star (e.g., van 
Kerkwijk et al.\ 1992; 1996). The nature of the compact object is uncertain; it 
may be either a neutron star or a black hole (e.g., Schmutz et al.\ 1996; Ergma 
\& Yungelson 1998; Stark \& Saia 2003). The system, located at a distance of 
$\sim$9 kpc (e.g., Predehl et al.\ 2000), is embedded in such a dense wind from 
the donor star that most of its intrinsic emission is strongly obscured. Still, 
its hard X-rays and \g-rays can be observed unimpeded, making it a very good 
target for \integral. 

As an X-ray source, Cyg X-3 is one of the brightest in the Galaxy. It displays 
states with high and low level of emission at $\lesssim$10 keV (e.g.\ White \& 
Holt 1982; Watanabe et al.\ 1994). The observed $EF_E$ spectrum in the high/soft 
state peaks at a few keV and is highly variable, whereas in the low/hard state 
the peak is at $\sim$20 keV and there is less variability. At the other end of 
the spectrum, Cyg X-3 is the strongest radio source among the X-ray binaries and 
shows huge radio outbursts associated with relativistic jets. The radio activity 
and the production of the jets are closely linked with the X-ray emission and 
the different X-ray states (McCollough et al.\ 1999; Choudhury et al.\ 2002). 

It is believed that Cyg X-3 shows so many unique properties because it is in a 
very short-lived transitional phase of binary evolution. It is therefore 
possible that the system represents a stage of evolution that will be the fate 
of many of the massive X-ray binaries observed today. If this is the case, an 
understanding of this peculiar source could prove to have broad implications for 
high-mass X-ray binaries in general.

\section{Observations and data analysis}

Cyg X-3 has been observed on several occasions with all three X/\g-ray 
instruments aboard \integral. The most extensive observations were carried out 
during the \integral\/ performance verification phase observations of the Cygnus 
region in December 2002. (The data for Rev.\ 23 are described in Vilhu et al.\ 2003, hereafter V03.) Cyg X-3 has also been in the FOV in several of the 
Galactic Plane Scans (GPS) carried out during 2003. Due to the large variability 
of the source (see Fig.\ 1), a simple co-adding of all available observations 
will not represent a physical spectrum of Cyg X-3. It is true, however, that the 
spectral state of this source can be roughly determined by the flux level in 
1.5--12 keV X-rays as measured by the All Sky Monitor (ASM) on board \xte. Here, we have therefore grouped the observations based on the \asm\/ 
count rate and combined the selected science windows to build two co-added 
spectra of Cyg X-3, representing different spectral states. The dates for the 
observations are listed in Table 1 and also indicated in Fig.\ 1 together with 
the corresponding flux levels as seen by \asm. 

\begin{figure*}
\begin{minipage}{0.5\textwidth}
\includegraphics[width=\linewidth]{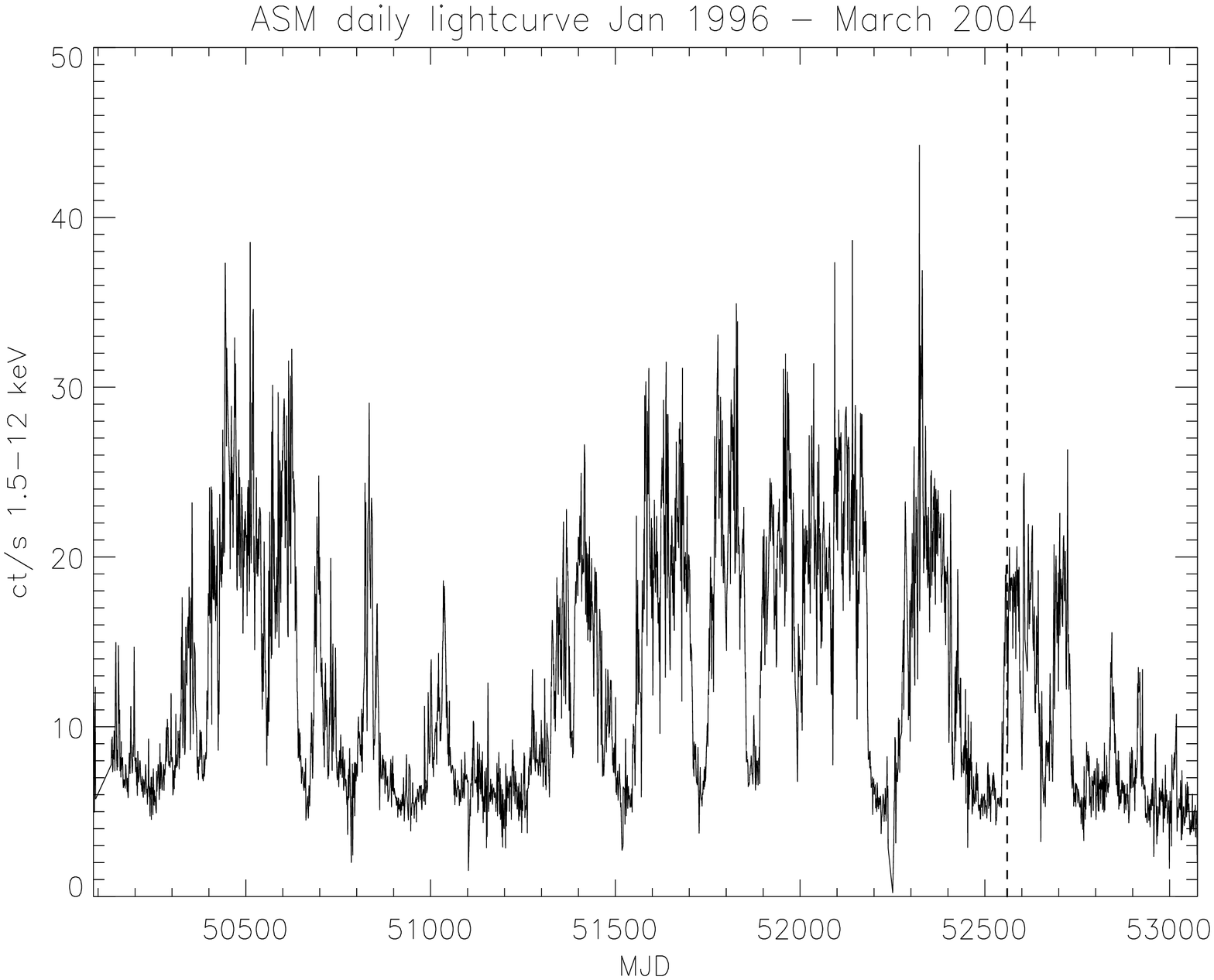}
\end{minipage}
\begin{minipage}{0.5\textwidth}
\includegraphics[width=\linewidth]{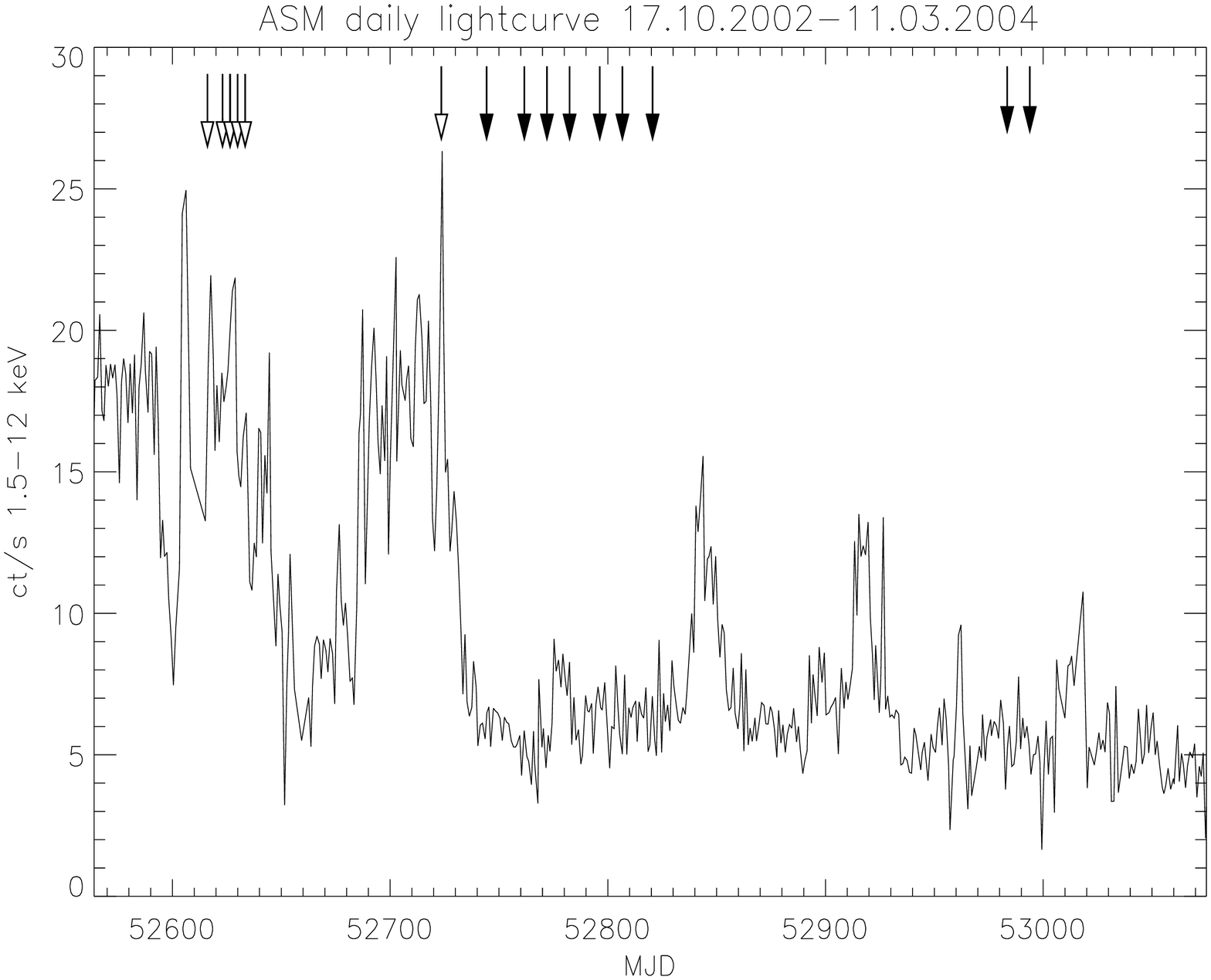}
\end{minipage}
\caption{Left panel: The \asm\/ lightcurve since launch of \rxte\/ in January 1996 until March 2004. The launch of \integral\ is shown by a vertical line. Right panel: An enlarged portion of the same lightcurve representing the lifetime of \integral. The times of the \integral\/ observations are shown as arrows. Observations indicated with open arrows were used to construct the `high' state spectrum and observations indicated by filled arrows to construct the `low' state` spectrum. It is clear from a comparison with the long-term \asm\/ lightcurve that the source has not been in a very high state since the launch of \integral. }
\label{rxtespectra}
    \end{figure*}

\begin{table}[h]
\caption{Data from the listed revolutions were used to create the `high' state spectrum. The dates represent the start date of each revolution. The same dates are indicated by open arrows in Fig. 1.\ Effective exposure times in ksec are given for each instrument per revolution and in total.}
\label{highrev}
$$
\begin{array}{lccccc}
            \hline
            \noalign{\smallskip}
{\rm Rev.} & {\rm start\ date} & {\rm JEM\!-\!X} & {\rm ISGRI} & {\rm SPI}\\
 \noalign{\smallskip}
          \hline
            \noalign{\smallskip}
19 & {\rm 2002-12-09} & 36 & 36 & -\\
20 & {\rm 2002-12-12} & 17 & 17 & 133\\
21 & {\rm 2002-12-15} & - & - & 180\\
22 & {\rm 2002-12-18} & - & - & 161\\
23 & {\rm 2002-12-21} & 82 & 82 & 169\\
54 & {\rm 2003-03-24} & 2 & 2 & 9\\
\hline
\noalign{\smallskip}
{\rm Total} & {\rm exposure\ time} & 137  & 137   & 652\\
\noalign{\smallskip}
\hline
         \end{array}
     $$ 
\end{table}

\begin{table}[h]
\caption{Data from the listed revolutions were used to create the `low' state spectrum. The dates represent the start date of each revolution. The same dates are indicated by filled arrows in Fig. 1. Effective exposure times in ksec are given for each instrument per revolution and in total.}
\label{}
$$
\begin{array}{lccccc}
            \hline
            \noalign{\smallskip}
{\rm Rev.} & {\rm start\ date} & {\rm JEM\!-\!X} & {\rm ISGRI} & {\rm SPI}\\
 \noalign{\smallskip}
          \hline
            \noalign{\smallskip}
59 & {\rm 2003-04-08} & 4 & 2 & -\\
62 & {\rm 2003-04-17} & - & - & 11\\
67 & {\rm 2003-05-01} & - & - & 12\\
70 & {\rm 2003-05-10} & 2 & 2 & 11\\
74 & {\rm 2003-05-22} & 4 & 4 & 11\\
79 & {\rm 2003-06-06} &  4 &  4 & 12\\
82 & {\rm 2003-06-15} & 2 & 2 & 11\\
87 & {\rm 2003-06-30} & 2 & 2 & 14\\
142 & {\rm 2003-12-12} & 2 & 2 & 11\\
145 & {\rm 2003-12-21} & 2 & 2 & 11\\
\hline
\noalign{\smallskip}
{\rm Total} & {\rm exposure\ time}: & 22 & 20 & 104\\
\noalign{\smallskip}
\hline
         \end{array}
     $$ 
\end{table}

\subsection{JEM-X}

The observations were performed with the second (JEM-X2) of the two identical 
X-ray monitors. The JEM-X and ISGRI data were analysed using ISDC's (Courvoisier 
et al.\ 2003) Offline Standard Analysis (OSA) software version 3.0. The data 
were taken from 54 pointings from the high state and 11 pointings in the low 
state with the source always in the partially coded field-of-view, within 
$5^\circ$ from the FOV center. Both Cyg X-1 and SAX J2103.5+4545 (an X-ray 
pulsar, discovered by Hulleman, in 't Zand \& Heise 1998) are $>$10$^\circ$ 
from the FOV center and hence the data are not contaminated by those sources. 
Source spectra were extracted individually per pointing and then added together 
into an average 'high' and 'low' spectrum respectively. The spectra were 
implicitly background-subtracted by a deconvolution algorithm assuming a 
spatially flat background. We used the energy range of 3.0--20 keV for spectral 
fitting. Systematic errors were added as 20\% for channels 1--58, 10\% for 
channels 59--96 and 2\% for channels 96--255.

\subsection{IBIS/ISGRI}

For ISGRI, we have selected 54 INTEGRAL pointings for the high state and 10 for 
the low state with the source always in the totally coded field-of-view. This way we have simultaneous ISGRI and 
JEM-X data and we do not use the ISGRI partially coded field-of-view for which the ISGRI 
response matrix is known to be less accurate. The two different spectral-state 
set of pointings have been analysed separately. For each set we have produced a 
mosaicked image (weighted combination of single pointing images) in 10 energy 
bands. We built the high and soft spectrum extracting the counts/sec from the 
brightest pixel in each energy band. This method of extracting the spectra from 
the mosaics instead of adding individual spectra significantly reduces the level 
of noise and  gives high-quality spectra even for low count rates. For the 
fitting of the two broad-band spectra, we used  the energy range 20--150 keV. A 
5$\%$ systematic error was added to all channels.

\subsection{SPI}

SPI observations within $15^\circ$ from the FOV center, from times with similar flux levels at 1.5--12 keV as for the JEM-X and ISGRI data, were selected. The SPI data were 
analysed using OSA version 3.0, except for some updates due to known problems in 
the binning software. For image reconstruction and spectral extraction we 
applied version 6.0 of the SPI Iterative Removal Of Sources program (SPIROS; 
Skinner \& Connell 2003). This allowed us to apply a new background model based 
on the mean count modulation of the detector array. In order to get precise 
fluxes, the source positions of the known sources in the field of view have been 
fixed to their catalogue values. In addition, we allowed SPIROS to apply time 
dependent normalisation to the source fluxes of the three brightest and variable 
sources (Cygnus X-1, Cygnus X-3, and EXO 2030+375). Thus the SPI spectrum of 
Cygnus X-3 presented here is an average over 10 ksec time bins. For spectral 
extraction, 25 logarithmic bins in the 20--100 keV energy range, and 5 
logarithmic bins in 100--2000 keV have been used. The energy range 20--200 keV 
was used for the spectral fitting due to strong background features at higher 
energies. The instrumental response function has been derived from on-the-ground 
calibration (Sturner et al.\ 2003) and then corrected based on the Crab 
calibration observation. A 5\% systematic error was added to all channels.

The resulting two combined JEM-X, ISGRI and SPI broad-band spectra were fitted using the XSPEC package (Arnaud 1996). The ISGRI and SPI spectra have been renormalized to the JEM-X data. The ISGRI and JEM-X levels showed good agreement while the SPI normalization is a factor of $\sim$2 higher. The obtained `high' spectra correspond to the 1.5--12 keV source flux $\gtrsim$200 mCrab, and the `low' spectra, to $\lesssim$130 mCrab. Note that neither of these states represent the extreme high/soft or low/hard states displayed by this source on a longer timescale (see \S 4 below).

\section{Broad-band spectral modelling}

\begin{table*}
\centering
\begin{footnotesize}
\caption{Model parameters$^{\mathrm{a}}$ for the two broad-band \integral\/ spectra of Cyg X-3 in the `high' and `low' state.}
\label{par}
$$   
\begin{array}{lcccccccccccc}
\hline
\noalign{\smallskip}
{\rm Data}&N_{\rm H,0}&N_{\rm H,1}&f_1&\lh/\ls&\tau&kT^{\mathrm{b}}& \Omega/ 2\pi& \xi^{\mathrm {c}}& \efe^{\mathrm{d}}& \ffe& {F_{\rm bol}}^{\rm e}& \chi^2/\nu \\
\noalign{\smallskip}
& 10^{22}\, {\rm cm}^{-2}&10^{22}\, {\rm cm}^{-2}&&&&{\rm keV}&&{\rm erg\, cm}\, {\rm s}^{-1}& {\rm keV}&{\rm cm}^{-2}\, {\rm s}^{-1}&{\rm erg\, cm}^{-2}\, {\rm s}^{-1}\\
 \hline
 \noalign{\smallskip}
{\rm High} & 20.7\!\pm\! 1.4 & 570 \!\pm\! 50 & 0.14\!\pm\! 0.11 & 0.21 \!\pm\! 0.03 & 0.21\!\pm\! 0.04 & 73 & 0.6\!\pm\! 0.4 & 7000_{-7000}^{+3000} & 6.61\!\pm\! 0.12 & 0.0085 & 5.2\!\times\! 10^{-9} & 129/172\\
\noalign{\smallskip}
{\rm Low} & 8.6\!\pm\! 7.3 & 32 \!\pm\! 21 & 0.75\!\pm\! 0.20  & 0.22\!\pm\! 0.06 & 0.15 \!\pm\! 0.13 & 102 & 2.0\!\pm\! 1.2 & 3000\!\pm\! 1000 &  6.87\!\pm\! 0.07 & 0.0146 & 3.8\!\times\! 10^{-9} & 129/147\\
\hline
 \noalign{\smallskip}
 \end{array}
     $$ 
\begin{list}{}{}
\item[$^{\mathrm{a}}$] The uncertainties are the linear uncertainties as given by XSPEC; the blackbody temperature was fixed at 0.38 keV.
\item[$^{\mathrm{b}}$] Calculated from the energy balance, i.e., not a free fit parameter. 
\item[$^{\mathrm{c}}$] Assumed $\leq 10^4$ in the fits.
\item[$^{\mathrm{d}}$] The line photon flux corrected for the Galactic $N_{\rm H}$ only.
\item[$^{\mathrm{e}}$] The bolometric flux of the {\it absorbed\/} model spectrum and normalized to the JEM-X/ISGRI data.
\end{list}
\end{footnotesize}
\end{table*}

In spite of a large number of X-ray observations, the spectra of Cyg X-3 have mostly been interpreted in terms of phenomenological models (e.g.\ Nakamura et al.\ 1993), such as power law and bremsstrahlung. The first physical interpretation of the intrinsic broad-band spectrum of Cyg X-3, based on simultaneous \integral\/ (Rev.\ 23) and \xte/PCA-HEXTE observations in 2002 Dec.\ 22--23, was given by V03. Their model assumes a hot Comptonizing plasma in the vicinity of an accretion disc, which both provides soft seed photons for the Comptonization and reflects the hard photons emitted by the hot plasma. The specific model, which we also use in this work, is {\tt eqpair} (Coppi 1992, 1999). 

In general, the electron distribution in the hot plasma can be purely thermal or 
hybrid, i.e., Maxwellian at low energies and non-thermal at high energies, if an 
acceleration process is present.  This distribution, including the electron 
temperature, $T$, is calculated self-consistently from the assumed form of the 
acceleration (if present) and from the luminosities corresponding to the plasma 
heating rate, $L_{\rm h}$, and to the seed photons irradiating the cloud, 
$L_{\rm s}$.  The plasma optical depth, $\tau$,  includes a contribution from 
e$^\pm$ pairs. The importance of pairs depends on the ratio of the luminosity to 
the characteristic size, $r$, which is usually expressed in dimensionless form 
as the compactness parameter, $\ell \equiv L\sigma_{\rm T}/(r m_{\rm e} c^3)$, 
where $\sigma_{\rm T}$ is the Thomson cross section and $m_{\rm e}$ is the 
electron mass. 

In the fitting process, we have found that our both broad-band \integral\/ 
spectra are compatible with the hot plasma being completely thermal, and $kT\ll 
511$ keV. Then, the \ee\ pair production is negligible, and the absolute value 
of the compactness is only weakly important. Accordingly, we assume a constant 
$\ls=10$ (as in V03). We stress, however, that our data, limited to energies 
$\lesssim$200--300 keV, rather weakly constrain possible non-thermal component 
of the electron distribution and the presence of pairs.  Non-thermal processes  
certainly exist in at least some states of Cyg X-3 (Szostek \& Zdziarski 2004; 
see \S 4 below).

\begin{figure*}
\centering
   \includegraphics[width=0.90\linewidth]{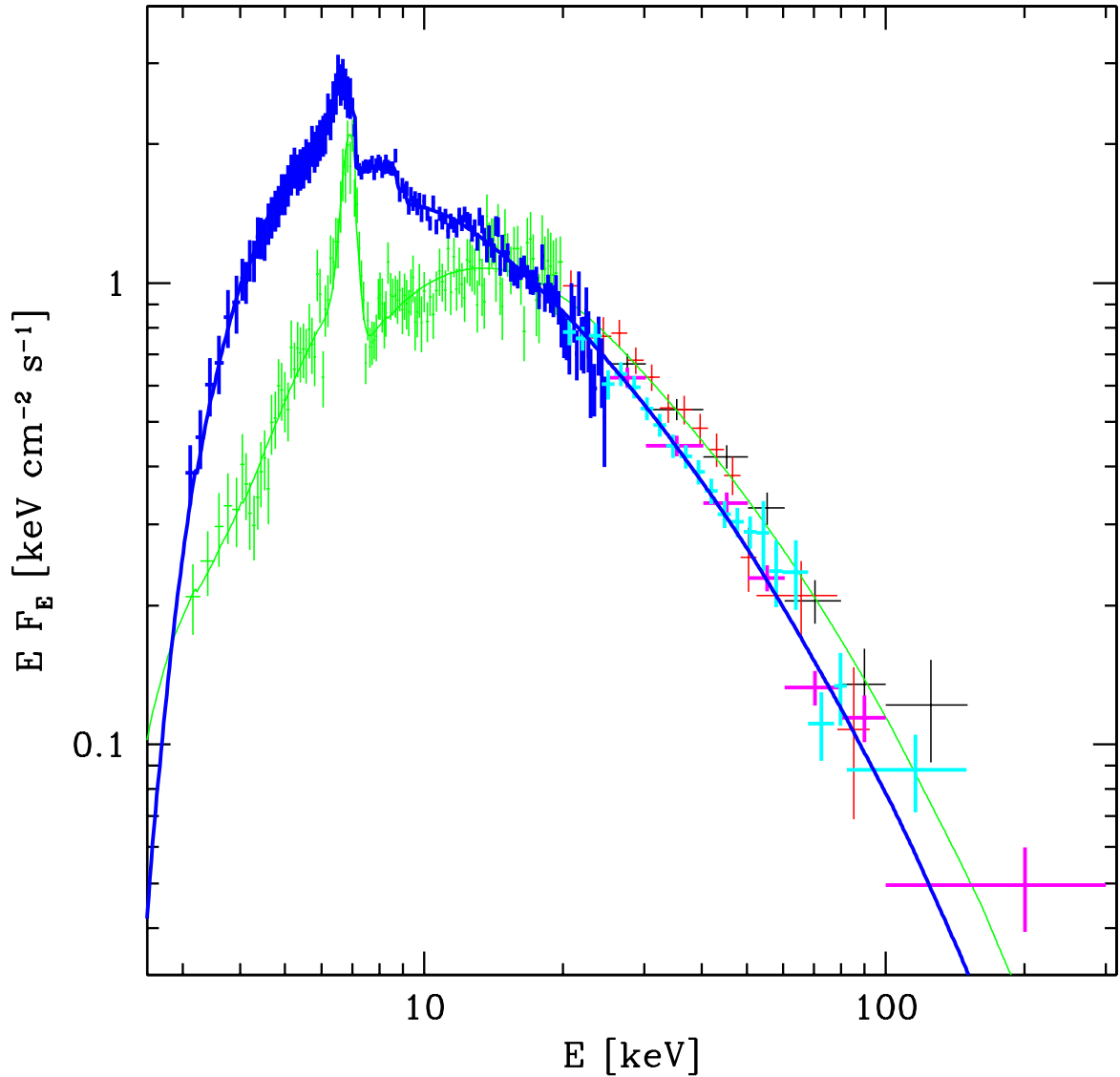}
\caption{Deconvoled spectra of Cyg X-3. The green, red and black data points correspond to the JEM-X, ISGRI and SPI spectra, respectively, in the high state. The heavy blue, cyan and magenta data points are for the JEM-X, ISGRI and SPI, respectively,  in the low state. The model spectra are shown in the green and blue curves. The ISGRI and SPI spectra have been renormalized to the JEM-X data. The proportions of this figure correspond to equal length per decade on each axis.}
\label{spectra}
\end{figure*}

Our data, covering only photon energies $\gtrsim$3 keV, rather poorly constrain 
the form of the seed photons. We therefore use here the same simple assumption 
as in V03, i.e, emission of an accretion disc with the maximum temperature fixed 
in the fit. The data also do not allow to constrain the structure of the X-ray 
absorber and the likely presence of additional spectral components in soft 
X-rays (e.g.\ Molnar \& Mauche 1986; Nakamura et al.\ 1993). This absorption and 
emission is due to the dense wind from the companion Wolf-Rayet star. These 
processes (which we do not treat in detail here) may strongly affect the spectra up to as much as $\sim$20 keV.

For the purpose of modelling the broad-band spectra, we use the same  
phenomenological model of the absorber as V03. It assumes that one absorbing 
medium with the column density, $N_{\rm H,0}$, fully covers the source, and 
another medium with the column, $N_{\rm H,1}$, covers a fraction, $f_1$, of the 
source. We also include Compton reflection (Magdziarz \& Zdziarski 1995), 
parametrized by an effective solid angle subtended by the reflector as seen from 
the hot plasma, $\Omega$, and assuming an inclination of $60^\circ$. We also 
include an Fe K$\alpha$ fluorescent line, which we model as a Gaussian (with the 
width of $\sfe$, the photon flux of $\ffe$, and the peak energy at $\efe$). We 
allow the reflecting medium (at the assumed temperature of $10^6$ K) to be 
ionized, using the ionization calculations of Done et al.\ (1992). The ionizing 
parameter is defined as $\xi\equiv 4 \pi F_{\rm ion}/n$, where $F_{\rm ion}$ is 
the ionizing flux and $n$ is the reflector density. Given the simplified 
treatment of the ionized reflection of Done et al.\ (1992), we assume $\xi\leq 
10^4$. 

The obtained model parameters are given in Table 3, and the deconvolved data and 
the model spectra are shown in Fig.\ 2. We see that the two spectra differ strongly at low energies, but their high-energy tails, at energies $\gtrsim$20 keV, are almost identical, with only a slight difference in the normalization. 

\section{Discussion}

Somewhat surprisingly, the two average spectra obtained by us are very similar at energies above 20 keV, even if there is a clear difference between the shape below 20 keV. We can compare them to the range of the spectra observed so far by the pointed instruments on board \xte. A study of the different spectral states observed by the PCA and HEXTE was performed by Szostek \& Zdziarski (2004), who used the same model as that used by us and by V03. Fig.\ 3 shows four out of five average spectra of Szostek \& Zdziarski (2004), with the second hardest spectrum (group \#2) omitted for clarity. Those spectra have also been shown by Zdziarski \& Gierli\'nski (2004) to be similar to those of the canonical states of black-hole binaries (notwithstanding the peculiar form of absorption in Cyg X-3).

In Fig.\ 3, we also show the observed `low' and `high'  \integral\/ spectra. We find them to be relatively similar in shape to the hard (blue) and intermediate (magenta) average \xte\/ spectra. However, the normalization of the \integral\/ spectra is lower by a factor of $\sim$2. A part of this normalization difference is likely to be due to the different absolute calibration of the JEM-X and ISGRI vs.\ that of the PCA. 

\begin{figure}
\centering
   \includegraphics[width=\linewidth]{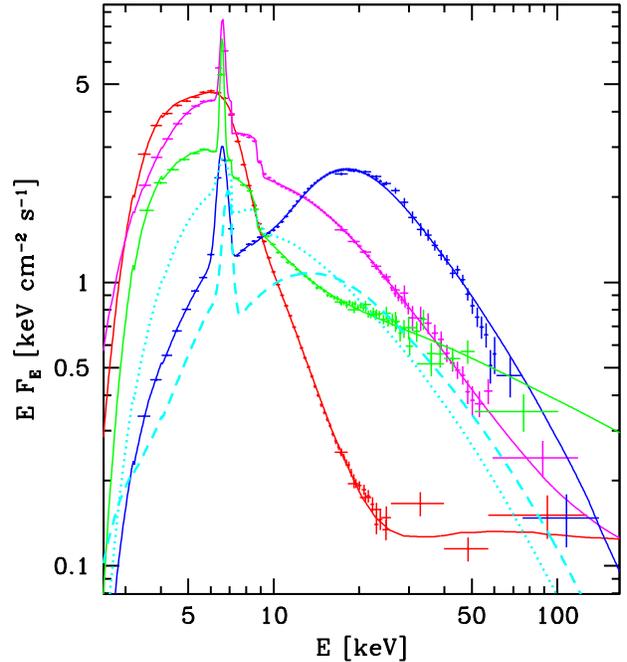}
\caption{The range of spectra observed so far by \xte/PCA/HEXTE compared to the present \integral\/ spectra. The red, magenta, green and blue spectra represent four out of five groups obtained by Szostek \& Zdziarski (2004) by averaging 
42 \xte\/ observations from 1996--2000 (normalized to the PCA). The low and high \integral\/ spectral models are plotted in the dashed and dotted cyan curves, respectively. }
\label{kyoto}
    \end{figure}

From Fig.\ 3, we can clearly conclude that with \integral, we have as yet 
sampled neither the hardest nor the softest spectra of Cyg X-3. Of particular 
interest here is the ultrasoft state (red spectrum in Fig.\ 3) with a distinct 
hard tail. The presence of such a high energy tail without a cutoff below 
$\sim$200 keV requires a nonthermal electron distribution and an acceleration 
mechanism operating. Due to the fact that the HEXTE detector can measure it only 
up to $\sim$100 keV, the detailed form of this hard tail in the high/ultrasoft 
state has remained largely unexplored, and it would be of great importance to 
observe it with \integral. The presence of that tail is clearly accompanied by 
the very highest level of the flux at 1.5--12 keV, which is also monitored by 
the \xte/ASM. All ultrasoft spectra corresponding to the average shown in Fig.\ 
3 were observed when the ASM flux level was $>$27 counts/sec, which corresponds 
to $>$360 mCrab. In comparison, see the right panel of Fig.\ 1, 
the highest flux levels observed during the lifetime of \integral\/ have only 
occasionally reached above 20 counts/sec and \integral\/ has thus yet not been 
able to observe a true high/ultrasoft state of Cyg X-3.

\section{Conclusions}

We have combined multiple JEM-X, ISGRI and SPI spectra from available public and 
GPS observations to create two broad-band spectra of Cyg X-3, corresponding to 
high and low levels of the 1.5--12 keV flux. The resulting `high' and `low' 
spectra resemble what can be referred to as the intermediate and the low/hard 
state of this source, but are similar in shape above 20 keV. We have fitted the 
intrinsic spectra with thermal Comptonization models including Compton reflection, 
which are (see Zdziarski \& Gierli\'nski 2004) similar to those found for 
black-hole binaries. We point out that \integral\/ has yet to observe Cyg X-3 in 
the high/ultrasoft state, which would provide a very interesting opportunity to 
study the hard non-thermal tail associated with this state. 

\section*{Acknowledgments}
This research has been based on observations with {\it INTEGRAL}, an ESA project with instruments and science data center funded by ESA and member states (especially the PI countries: Denmark, France, Germany, Italy, Switzerland, and Spain), the Czech Republic, and Poland and with the participation of Russia and the US. It has also made use of data obtained through the HEASARC Online Service (provided by NASA/GSFC), and of NASA's Astrophysics Data System. The authors from the Observatory of the University of Helsinki acknowledge the Academy of Finland, TEKES, and the Finnish space research programme ANTARES for financial support. AAZ has been supported by KBN grants 5P03D00821, 2P03C00619p1,2 and PBZ-KBN-054/P03/2001.

\end{document}